\documentclass[journal]{IEEEtran}

\usepackage[utf8]{inputenc} 
\usepackage[T1]{fontenc}    
\usepackage{nicefrac}       
\usepackage{microtype}      

\usepackage{color}
\usepackage{xcolor}
\definecolor{myblue}{rgb}{0.0, 0.0, 0.5}

\usepackage[pagebackref=false,colorlinks=true,bookmarks=false,linkcolor=myblue,citecolor=myblue,urlcolor=myblue]{hyperref}
\usepackage{url}            
\usepackage{booktabs}
\usepackage{arydshln} 
\usepackage{multirow}
\usepackage{graphicx}
\usepackage{wrapfig}
\usepackage{subcaption}
\usepackage{comment}

\usepackage{amssymb, bbm} 

\usepackage{amsmath,amsfonts,bm}

\def\eqref#1{equation~\ref{#1}}

\def\ceil#1{\lceil #1 \rceil}

\def\1{\bm{1}}

\DeclareMathAlphabet{\mathsfit}{\encodingdefault}{\sfdefault}{m}{sl}
\SetMathAlphabet{\mathsfit}{bold}{\encodingdefault}{\sfdefault}{bx}{n}

\def\gF{{\mathcal{F}}}

\def\gU{{\mathcal{U}}}

\def\sR{{\mathbb{R}}}

\newcommand{\E}{\mathbb{E}}

\DeclareMathOperator*{\argmax}{arg\,max}

\usepackage{cite}

\usepackage{algorithm}
\usepackage[noend]{algpseudocode}
\usepackage[font=normalsize,labelfont=bf]{caption}

\newcommand{\etal}{\emph{et al.~}}

\begin{document}

\title{RAIN: A Simple Framework for Robust and Accurate Image Classification}

\author{
{Jiawei Du*, Hanshu Yan*, Vincent Y. F. Tan,~\IEEEmembership{Senior~Member,~IEEE,}
Joey Tianyi Zhou, Rick Siow Mong Goh, Jiashi Feng~\IEEEmembership{Member,~IEEE}}
\thanks{Jiawei Du is with the Department of Electrical and Computer Engineering, National University of Singapore and the Institute of High Performance Computing, A*STAR, Singapore.}
\thanks{Hanshu Yan, Vincent Y. F. Tan, and Jiashi Feng are with the Department of Electrical and Computer Engineering, National University of Singapore.}
\thanks{Joey Tianyi Zhou and Rick Siow Mong Goh are with the Institute of High Performance Computing, A*STAR, Singapore.}
\thanks{*~Equal Contribution, e-mail: hanshu.yan@u.nus.edu.}
}

\maketitle

\begin{abstract}
It has been shown that the majority of existing adversarial defense methods achieve robustness at the cost of sacrificing prediction accuracy. The undesirable severe drop in accuracy adversely affects the reliability of machine learning algorithms and prohibits their deployment in realistic applications. This paper aims to address this dilemma by proposing a novel preprocessing framework, which we term \emph{\underline{R}obust and \underline{A}ccurate \underline{I}mage classificatio\underline{N}}~(RAIN), to improve the robustness of given CNN classifiers and, at the same time, preserve their high prediction accuracies. RAIN introduces a new randomization-enhancement scheme. It applies randomization over inputs to break the ties between the model forward prediction path and the backward gradient path, thus improving the model robustness. However, similar to existing preprocessing-based methods, the randomized process will degrade the prediction accuracy. To understand why this is the case, we compare the difference between original and processed images, and find it is the loss of high-frequency components in the input image that leads to accuracy drop of the classifier. Based on this finding, RAIN enhances the input's high-frequency details to retain the CNN's high prediction accuracy. Concretely, RAIN consists of two novel randomization modules: randomized small circular shift (RdmSCS) and randomized down-upsampling (RdmDU). The \emph{RdmDU} module randomly downsamples the input image, and then the \emph{RdmSCS} module circularly shifts the input image along a randomly chosen direction by a small but random number of pixels. Finally, the RdmDU module performs upsampling with a detail-enhancement model, such as deep super-resolution networks. We conduct extensive experiments on the STL10 and ImageNet datasets to verify the effectiveness of RAIN against various types of adversarial attacks. Our numerical results show that RAIN outperforms several state-of-the-art methods in both robustness and prediction accuracy. The proposed RAIN framework works in a plug-in manner and does not require to modify the given CNNs, thus easy and efficient for implementation.\footnote{The source codes of RAIN project are available at \url{https://github.com/dydjw9/RAIN}.}

\end{abstract}


%
\IEEEpeerreviewmaketitle

\section{Introduction} \label{introduction}
\IEEEPARstart{I}{n} recent years, convolution neural network (CNN)-based image classification models  have been successfully applied to a variety of areas such as robot vision \cite{giusti2015machine}, biometrics authentication \cite{sun2014deep}, and autonomous vehicles \cite{papernot2016practical}. In these real-world applications, the CNN models not only should make accurate predictions, but also need  to be robust against small perturbations over input, i.e., the models should make consistent decisions for examples that have been contaminated by a small amount of noise.  However, CNN classifiers are known to be highly vulnerable to adversarial examples \cite{szegedy2013intriguing,biggio2013evasion}\textemdash even certain visually imperceptible perturbations to inputs can easily fool the CNNs, resulting in grossly incorrect predictions. To ameliorate this problem, researchers have developed a variety of adversarial defense methods to protect CNN classifiers from adversarial attacks. These methods can be roughly divided into two categories: adversarial training \cite{madry2017towards, xie2019feature,zhang2020attacks} and input preprocessing \cite{xie2017mitigating, dziugaite2016study, mustafa2019image}.

The adversarial training (AT)-based methods \cite{madry2017towards, xie2019feature,zhang2019you,mosbach2018logit} train the CNN classifiers on the clean images as well as their adversarial versions generated on-the-fly during the model training procedure. Although they can effectively improve the model robustness, generating adversarial examples via gradient descent per training iteration is highly computationally demanding. Thus, the implementation of AT-based methods is inefficient in practice. Unlike the AT-based methods, the input preprocessing methods develop randomized  or non-differentiable transformations to block the path of gradient backpropagation from the prediction to input, so that the attackers cannot access inner gradients to craft adversarial examples, thus, fail to fool the classifiers. These methods do not need to re-train the classifiers, and thus are more computationally efficient \cite{song2017pixeldefend,xie2017mitigating,prakash2018deflecting, samangouei2018defense,mustafa2019image,kou2020enhancing}. 
In real-world applications, 
classifiers are usually uninformed about whether each input is clean or adversarially perturbed. Thus, to ensure reliability, the classifiers have to treat clean and perturbed images on the same footing and should make correct predictions for both types of images. However, existing methods in both of these two categories suffer from the severe   classification accuracy drop on clean inputs when improving the robustness of CNN classifiers \cite{zhang2019theoretically, xie2017mitigating}. Mitigating the trade-off between accuracy and robustness is very challenging; existing works that target this problem are still few and far between. 

In this work, we follow the strategy of input processing and aim to develop an efficient and effective defense framework, which also can alleviate the problem of the drop in accuracy.
Our methodology is surprisingly straightforward: On the one hand, since adversarial attack methods mostly leverage the gradients with respect to the input to generate adversarial examples, to defend against adversarial examples, we can use randomization to decorrelate the inference forward path and the gradient backward path \cite{xie2017mitigating}. On the other hand, to maintain high prediction accuracy, we should utilize image transformations to which CNN classifiers are (almost) invariant. 

Taking into account these two aspects, we develop two novel randomized preprocessing modules---the randomized small circular shift (RdmSCS) and the randomized down-upsampling (RdmDU). In the RdmSCS module, input images are circularly shifted along a randomly chosen direction by a small and random number of pixels. Since CNNs are shown to be invariant to a small shift \cite{KaudererAbrams2018QuantifyingTI}, the proposed RdmSCS will not degrade the natural accuracy by much. To further improve the robustness, we propose the RdmDU module, in which the input images are first downsampled through randomly sampling one pixel from each corresponding small patch. We then resize the low-resolution image into the original size with certain upsampling method. If using the widely-used bicubic interpolation for upsampling, RdmDU would also incur a clear drop in accuracy as the existing preprocessing-based methods do. To understand why the accuracy degrades, we conduct an empirical study which reveals that the loss of high-frequency details is a critical reason for the drop in accuracy. Based on this observation, we choose a detail-enhancement model, such as a deep super-resolution (SR) network, to reconstruct (i.e., obtain upsampled) high-quality images with rich (i.e., high-frequency) details. Consequently, the detail-enhancement RdmDU module manages to preserve high natural accuracy.

Combining these two randomized modules, we propose the \emph{\underline{R}obust and \underline{A}ccurate  \underline{I}mage classificatio\underline{N}} (RAIN) framework. The input image is processed sequentially as follows: (i) Randomized downsampling, (ii) RdmSCS, (iii) SR upsampling. We conduct extensive sets of experiments on the STL10 \cite{coates2011analysis} and ImageNet \cite{deng2009imagenet} datasets to verify the effectiveness of the proposed RAIN. The results demonstrate that RAIN significantly enhances the robustness of given CNN classifiers (e.g., from 0\% to 52.3\% against C\&W attack on the ImageNet) and does not lead to a severe drop in the natural accuracy (e.g., from 100\% to 93.3\% on ImageNet). In summary, the main contributions of this work are as follows:
\begin{itemize}
  \item We propose the RAIN framework, which enhances the robustness and retains the high prediction accuracy of a given CNN classifier. This method achieves better or almost equal robustness against adversarial attacks and clearly better accuracy on clean images when compared to several state-of-the-art defense approaches \cite{xie2017mitigating,prakash2018deflecting,madry2017towards,xie2019feature}. 
  \item 
  We introduce two simple yet effective components within the RAIN framework: RdmSCS and RdmDU. Each of these modules can independently improve the robustness with a minor degradation to the prediction accuracy. 
  \item We reveal that the loss or inadvertent removal of high-frequency components in input images is a critical factor in the drop of the accuracy of a CNN classifier. Hence, we are inspired to use a detail-enhancement model for upsampling within the RdmDU module.
  \item RAIN works in a plug-in manner and is easy to implement on given any CNN classifier. 
  \end{itemize} 

The rest of this paper is organized as follows. In Section \ref{sec:related_works}, we review related existing adversarial defense methods. In Section \ref{sec:rain}, we first elaborate on the two dedicated randomized preprocessing modules, namely RdmSCS and RdmDU. Then, we explain why the use of a detail-enhancement model in the RdmDU helps to preserve good classification accuracy of the given CNN. Finally, we formally introduce the complete RAIN framework. In Section \ref{sec:exp}, we conduct extensive experiments to evaluate the ability of RAIN to robustify CNN classifiers and to preserve high natural accuracies for them. In Section \ref{sec:ablation}, we further study how the scale of shifting in the RdmSCS and the order of combination of two modules affect RAIN's performance. We also illustrate how the RAIN framework purifies the feature maps of adversarial examples. We briefly summarize our work in Section \ref{sec:conclusion} and discuss several topics for future research.


\section{Related Work} \label{sec:related_works}
Robustness measures the sensitivity of the accuracy of a classification model to perturbations of the inputs. Many approaches have been proposed to estimate the most harmful adversarial examples within a given neighborhood, such as FGSM \cite{goodfellow2014explaining}, DeepFool \cite{moosavi2016deepfool}, and PGD \cite{madry2017towards}. These methods compute adversarial perturbations based on available gradients evaluated at neighborhoods of the given input images. They can access the gradients in each layer and are thus called \emph{white-box} attacks. In contrast, several other methods \cite{chen2017zoo,ilyas2018black,tu2019autozoom,du2019query} propose to estimate the adversarial perturbation by using a number of queries. These approaches, without information about the gradients in each layer, are called \emph{black-box} attacks. In this work, we use five white-box and two black-box attack methods to evaluate the robustness of classifiers. 

To enhance the adversarial robustness of given classifiers, researchers have recently proposed a wide variety of defense methods. Here, we review two main classes of such methods, namely adversarial training and input preprocessing methods. We also discuss the connections of these methods to our work.

\subsubsection{Adversarial Training} 
Adversarial training-based methods train CNN classifiers from scratch on both clean images and their corresponding adversarial examples to enhance their robustness. In each training epoch, the adversarial perturbations corresponding to the clean images are generated in real-time. The works in  \cite{madry2017towards, athalye2018obfuscated} show that adversarial training is generally an effective defense mechanism against certain types of adversarial perturbations. However, training CNNs from scratch and computing adversarial examples in real-time has an adverse effect on the training cost; in fact, a study \cite{schmidt2018adversarially} has shown that it increases the training cost by 3-to-30 times. 
This computational burden restricts the applicability of adversarial training methods on large-scale datasets. To ameliorate the problem, researchers have proposed several methods for efficiently estimating adversarial perturbations \cite{zhang2019you,shafahi2019adversarial}. Besides, another severe drawback of adversarial training methods is that the resultant classifiers tend to over-fit a certain class of adversarial examples \cite{rice2020overfitting,balaji2019instance,schmidt2018adversarially}. This results in a sharp drop in the prediction accuracy, and the classifiers are still vulnerable to other types of attacks. 

\subsubsection{Input Preprocessing}
Preprocessing-based defense methods remove the adversarial information from input images by transforming images in a dedicated way before they are fed into given CNN classifiers. One straightforward idea is to map input images onto the manifold generated by the clean data. To this end, several works \cite{samangouei2018defense, song2017pixeldefend} propose to train generative models on the original dataset and use them to reconstruct images that are close to the adversarial examples for making predictions. Similarly, the work of \cite{yang2019me} first impairs small patches of the adversarial examples and then performs reconstruction via matrix estimation. Traditional image processing operations have also been shown to be effective in robustifying given CNN classifiers. For example, Mustafa {\em et al.} \cite{mustafa2019image} propose to use wavelet denoising and SR techniques to deactivate the adversarial perturbations. 
However, the enhanced robustness that results from these methods is due to the fact that the attackers are not provided with the internal gradients of the preprocessing modules \cite{athalye2018obfuscated, samangouei2018defense}. Athalye \etal \cite{athalye2018obfuscated} proposes the Backward Pass Differentiable Approximation technique and can successfully bypass this problem. 
The works of \cite{xie2017mitigating,prakash2018deflecting} propose to randomize image processing operations such as padding and resizing. The randomization introduced effectively breaks the connection between the forward inference path and backward gradient path, and consequently, it improves robustness against adversaries. However, these randomization-based methods also suffer from deteriorations in prediction accuracies because CNN models are not invariant to the random padding and resizing operations. Athalye \etal \cite{athalye2018obfuscated} also demonstrates that the Expectation over Transformation technique is able to find the worst perturbation that can successfully mislead the CNN classifiers. 

Our work develops an effective and efficient preprocessing-based defense framework, which amalgamates randomization concepts into two image processing modules that preserve the overall classification accuracy. RAIN is easy to implement and can defend EoT-based attacks if combined with adversarially trained CNN classifiers. The work that is most related to RAIN is presented in \cite{mustafa2019image}. The difference on the usage of SR models is that the authors of \cite{mustafa2019image} upsample input images with SR networks without performing downsampling. This will leads to a severe performance drop when the upsampling factor becomes large (e.g., 3 or 4) because the given CNN classifier is trained with images that are of the original size and vanilla CNNs are not scale-invariant \cite{xu2014scale}.

\section{RAIN: Robust and Accurate Image Classification Networks}
\label{sec:rain}

In this section, we first introduce the background on robustness evaluation and formally define the problem to solve. Then, we elaborate the proposed randomization-enhancement scheme that contains two novel preprocessing modules. Finally, we present the full RAIN framework and its implementation details.

\subsection{Preliminaries}
\label{sec:preliminaries}
Given a CNN classifier $C(\cdot)$ and a perturbation budget $\epsilon$, the ideal untargeted adversarial example $x^{\text{adv}}_{\epsilon}$ of an input $x$ is defined as the data point in its $\epsilon$-neighborhood that maximizes certain classification loss. Formally,
\begin{equation}
    x^{\text{adv}}_{\epsilon} = \argmax_{x'\in \mathcal{B}(x,\epsilon)} ~ l(C(x'), y),
    \label{eq:adv_example}
\end{equation}
where $l(\cdot,\cdot)$ is the loss function, $y$ is the label of $x$, and $\mathcal{B}(x,\epsilon)$ is the $\epsilon$ ball around $x$, which defined based on certain norm, i.e., $\mathcal B(x,\epsilon)=\{x':\|x'-x\|\leq \epsilon \}$. 
Most of existing attack methods \cite{goodfellow2014explaining,madry2017towards} are based on the gradient descent to solve the optimization problem in Equation (\ref{eq:adv_example}). In the following, we use $A_{\epsilon}(\cdot,\cdot)$ to represent certain adversarial attack method and the ideal adversarial example $x^{\text{adv}}_{\epsilon}$ is approximated by $ A_{\epsilon}(C,x)$. 

The robustness of the given CNN $C(\cdot)$ can be quantified by the classification accuracy over the adversarial examples of input images. Since it is less meaningful to attack inputs that are already classified incorrectly, we follow the principle in \cite{xie2017mitigating, prakash2018deflecting} and randomly collect a set of labelled images $\mathcal{T}=\{(x_j,y_j)\}_{j=1}^M$ for robustness evaluation, where  $(x_j,y_j)$ denotes the $j$th sample   \emph{correctly classified} by $C(\cdot)$ before being attacked, i.e., $C(x_j) = y_j$. For a particular adversarial attack method $A_{\epsilon}(\cdot, \cdot)$, the robustness of $C(\cdot)$ is evaluated as
\begin{equation}
    R_{\mathcal{T}}(C, A_{\epsilon}) = \frac{1}{M}\sum^M_{j=1}\mathbbm{1}\left[C(A_{\epsilon}(C, x_j))=y_j\right].
    \label{eq:robustness_metric}
\end{equation}
To enhance the robustness of CNN classifiers, many defense methods have been proposed. 
Let $C'(\cdot)$ be the robustified version of $C(\cdot)$. The natural accuracy of $C'(\cdot)$ is quantified by the classification accuracy on the collected set $\mathcal T$, i.e.,
\begin{equation}
    P_{\mathcal{T}}(C') = \frac{1}{M}\sum^M_{j=1}\mathbbm{1}\left[C'(x_j)=y_j\right].
\end{equation}

Although most of the existing denfese methods manage to robustify given classifiers, they suffer from severe drops in accuracy \cite{zhang2019theoretically, dziugaite2016study}. For example, on the ImageNet, the Pixel Deflection method improves the robustness of the given CNN from 0 to 32.0\%, but the natural accuracy on $\mathcal T$ drops from 100\% to 85.8\% (refer to Table \ref{tab:whitebox_result}). To ensure the reliability of classification in real applications, the CNN classifiers have to make correct predictions for both of the clean images and their perturbed examples.
In this work, we aim to develop a preprocessing-based method to mitigate this trade-off, i.e., for the robustified model $C'$, $R_{\mathcal T}(C', A_{\epsilon})$ is siginicantly higher than $R_{\mathcal{T}}(C, A_{\epsilon})$ and the accuracy of $C'(\cdot)$ over $\mathcal T$ should remain above a high value.

\subsection{Randomized Input Preprocessing}
\label{sec:randomization}
 
Inspired by the invariance of CNNs to slight shifts and scalings \cite{KaudererAbrams2018QuantifyingTI, Kayhan2020OnTI}, we aim to utilize these properties to design robust and accurate CNN models. In particular, we consider two randomized transformation modules: RdmSCS and RdmDU. One the one hand, randomization can result in a mismatch between the forward prediction path and the gradient back-propagation path \cite{xie2017mitigating}. Hence, these two transformations are expected to be able to improve the adversarial robustness of classifiers. On the other hand, CNNs are capable of learning translation-invariant representations due to the gradual increase in the size of the receptive field \cite{Gens2014DeepSN} and the pooling operations \cite{Jaderberg2015SpatialTN}. The downsampling and subsequent upsampling operations also do not change the size of the images and the positions of the contents. Thus, the two transformations of interest would likely not degrade the prediction accuracy significantly.

\def \SubFigWidth {0.3} 
\def \SubImgWidth {0.95}
\begin{figure}[t!]
    \centering
	\begin{subfigure}{\SubFigWidth\linewidth}
		\centering
    	\includegraphics[width=\SubImgWidth \linewidth]{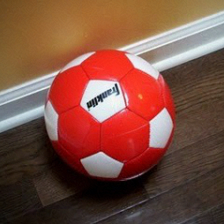}
    	\caption{Original}
	\end{subfigure}
	\begin{subfigure}{\SubFigWidth\linewidth}
		\centering
    	\includegraphics[width=\SubImgWidth \linewidth]{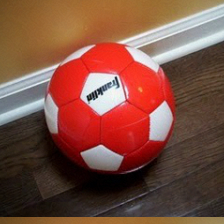}
    	\caption{$p=0.05$}
	\end{subfigure}
	\begin{subfigure}{\SubFigWidth\linewidth}
		\centering
    	\includegraphics[width=\SubImgWidth \linewidth]{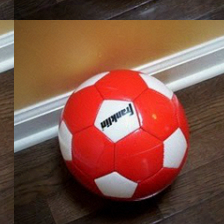}
    	\caption{$p=0.25$}
	\end{subfigure}
    \caption{An illustration of the RdmSCS module.  Left: an input image; Mid: the input shifted with $p=0.05$; Right: the input shifted with $p=0.25$.}
    \label{fig:shift_compare}
\end{figure}

\subsubsection{Randomized Small Circular Shift}
\label{sec:rdmscs}

We first consider a random shift operation of the input. Before it is fed to a given CNN model, the input is shifted by a random number of pixels along a randomly selected direction (horizontal or vertical). More precisely, two shifting parameters, $\Delta h$ and $\Delta w$, are randomly sampled from uniform distributions, i.e., $\Delta h \sim \gU(-hp, hp), \Delta w \sim \gU(-wp, wp)$, where $h$ and $w$ are the height and width of the input image respectively, and $p$ is a predefined positive constant. The sign of $\Delta h$ (or $\Delta w$) represents the vertical (or horizontal) direction for shifting, while the magnitude quantifies the number of pixels to be shifted. Algorithm \ref{alg:RdmSCS} elaborates on the implementation of the proposed random shifting operation. We use the upperscript $s$ to denote the shifted version of an image. 

\begin{algorithm}[h]
\caption{Randomized Small Circular Shift}
\label{alg:RdmSCS}
\begin{algorithmic}[1]
\Function {RdmSCS}{$x,p$}
    \For{$i_s$ in (0,$h-1$) and $j_s$ in (0,$w-1$)} 
        \State Randomly initialize  
            \Statex ~~~~~~~~~~ $\Delta h \sim \gU(-hp, hp)$, $\Delta w \sim \gU(-wp, wp)$
        \State $i\equiv(i^s+\Delta h) \mod h$
        \State $j\equiv(j^s+\Delta w) \mod w$
        \State $x^s(i^s,j^s)= x(i,j)$
    \EndFor
   \Return  $x^s$ 
\EndFunction
\end{algorithmic}
\end{algorithm}

Conventional translation operations pad zeros into the empty regions \cite{azulay2018deep}. Instead, our proposed operation splices back the shifted part of the input image circularly from the opposite boundary (see Figure \ref{fig:shift_compare}). The aim of this operation is to retain the frequency components and other statistics (such as the means and variances of pixel values) of original images as much as possible so that the preprocessing operation would not have an obvious negative impact on the prediction accuracy. For this reason, we usually set $p$ to be a small value in practice. Accordingly, we dub the proposed shifting operation as \emph{Randomized Small Circular Shift} (RdmSCS).

\begin{table}[h]
\caption{Evaluation of RdmSCS on the robustness and prediction accuracy on STL10 dataset (Refer to Section \ref{sec:exp_setting} for the setting of attack methods).}
    \centering
    \scalebox{1}{
      \begin{tabular}{c|c|cccc}
           & \textbf{Accuracy} & \multicolumn{3}{c}{ \textbf{Robustness}} \\
    \midrule
    \textbf{STL10} &   Clean Images & FGSM& C\&W&Deep Fool \\
            \hline
        CNN-only    & 1.000     & 0.074      &0.000      & 0.000\\        
	    $p = 0.01$	& 0.975     & 0.075	    & 0.117     & 0.000 \\
        $p=0.05$	& 0.958     & 0.215     &	0.421	& 0.093 \\
        $p=0.1$	    & 0.925	    & 0.311	    & 0.523	    & 0.252 \\
        $p=0.15$	& 0.836     & 0.299     &	0.671	& 0.422
    \end{tabular}
    }
    \label{tab:RdmSCS}
\end{table}

To understand how the value of $p$ affects the efficacy of RdmSCS, we evaluate the performance of the given CNN classifier equipped with RdmSCS. From Table \ref{tab:RdmSCS}, we can see that the proposed RdmSCS clearly improves on the robustness of CNN classifiers when $p$ is greater or equal to $0.05$. Choosing a large value of $p$ (e.g., $0.15$) can significantly robustify the CNN classifiers, but will degrade the accuracy on 
clean images because CNNs are not invariant to a large degree of shifting. To achieve a good balance between accuracy and robustness, the value of $p$ should be relatively small (less or equal to 0.1), so that the natural accuracy can be retained at a high value (above 90\%).

\subsubsection{Randomized Down-Upsampling} \label{sec:rdmdu}
In addition to the RdmSCS operation, we also introduce a \emph{Randomized Down-Upsampling} (RdmDU) module to further improve the robustness which simultaneously preserves the classification accuracy. Downsampling and subsequent upsampling do not change the size the input image and its semantic contents, thus, they are expected to maintain a good classification performance of the given CNN over clean images.  The RdmDU module consists of a randomized downsampling operation followed by an upsampling operation. Given an input image $x \in \sR^{h\times w \times 3}$, the random downsampling operation of RdmDU partitions the input into non-overlapping $2\times 2 $ patches and randomly picks one pixel from the four in each patch with the same probability $\nicefrac{1}{4}$. Consequently, the resultant image, denoted as $x^{\downarrow}$, has size of $  \frac{1}{2}h \times \frac{1}{2}w \times 3 $. As the CNN classifier is trained with images of the original size, the upsampling operation of RdmDU reconstructs a high-resolution image of size $h\times w\times 3$, denoted as $x^{\downarrow \uparrow}$, from $x^{\downarrow}$. Algorithm \ref{alg:RdmDU} shows the implementation of the RdmDU module in detail.

\begin{algorithm}[h]
\caption{Randomized Down-Upsampling}
\label{alg:RdmDU} 
\begin{algorithmic}[1]
\Require Input image $x$;
	\State $x=\Call{Random\_Downsampling}{x}$
	\State $x=\mathrm{Upsampling}(x,\mathrm{factor}=2)$
\Ensure $x$
\Function {Random\_Downsampling}{$x$} 
	\For{$i$ in $(0,\ceil{\frac{1}{2} h}-1)$ and $j$ in $(0,\ceil{ \frac{1}{2} w}-1 )$} 
	\State Randomly initialize 
	    \Statex ~~~~~~~~~~~~$\Delta i \sim \gU([0,1]),\Delta j \sim \gU([0,1])$
	\State $x^{\downarrow}(i,j) = x(\min(2i+{\Delta i},h-1),$
	    \Statex \hspace{1\algorithmicindent}  $\min(2j+{\Delta j},w-1))$
	\EndFor
   \Return  $x^{\downarrow}$ 
\EndFunction
\end{algorithmic}
\end{algorithm}

\def \SubFigWidth {0.16} 
\def \SubImgWidth {0.9}
\begin{figure*}[t!]
	\centering
	\begin{subfigure}{\SubFigWidth \linewidth}
		\centering
    	\includegraphics[width=\SubImgWidth \linewidth]{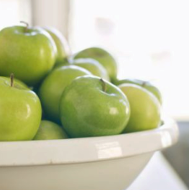}
    	\caption{$x$}
	\end{subfigure}
	\begin{subfigure}{\SubFigWidth \linewidth}
		\centering
    	\includegraphics[width=\SubImgWidth \linewidth]{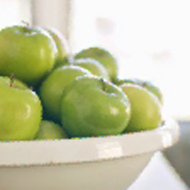}
    	\caption{$x^{\downarrow \uparrow}$}
	\end{subfigure}
	\begin{subfigure}{\SubFigWidth \linewidth}
		\centering
    	\includegraphics[width=\SubImgWidth \linewidth]{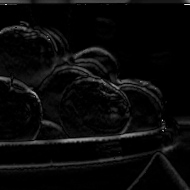}
    	\caption{$x-x^{\downarrow \uparrow}$}
	\end{subfigure}
	\begin{subfigure}{\SubFigWidth \linewidth}
		\centering
    	\includegraphics[width=\SubImgWidth \linewidth]{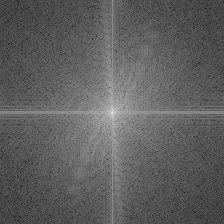}
    	\caption{$\mathcal{F}(x)$}
	\end{subfigure}
	\begin{subfigure}{\SubFigWidth \linewidth}
		\centering
    	\includegraphics[width=\SubImgWidth \linewidth]{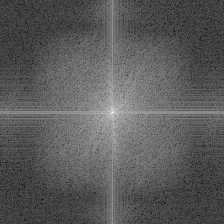}
    	\caption{$\mathcal{F}(x^{\downarrow \uparrow})$}
	\end{subfigure}
	\begin{subfigure}{\SubFigWidth \linewidth}
		\centering
    	\includegraphics[width=\SubImgWidth \linewidth]{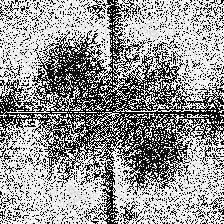}
    	\caption{$\mathcal{F}(x) - \mathcal{F}(x^{\downarrow \uparrow})$}
	\end{subfigure}
    \caption{Comparison between an input and the output of RdmDU-bic. From left to right: input $x$, output $x^{\downarrow \uparrow}$, difference $x-x^{\downarrow \uparrow}$; spectrum $\gF(x)$, spectrum $\gF(x^{\downarrow \uparrow})$, and $\gF(x)-\gF(x^{\downarrow \uparrow})$.}
    \label{fig:bic_compare}
\end{figure*}

For the upsampling operation in the RdmDU module, we first consider the bicubic interpolation method \cite{Keys1981CubicCI}, which is efficient and widely used in image processing tasks. We conduct experiments on the STL10 dataset to evaluate the effectiveness of the RdmDU-Bicubic module. From Table \ref{tab:RdmDU}, we find that both of the RdmDU-Bicubic can enhance the robustness of the CNN classifier against various kinds of attacks. However, the prediction accuracy of RdmDU-Bicubic modification drops significantly to 81.2\%, which is undesirable in high-stakes applications, such as the autonomous vehicles. To understand why upsampling with bicubic interpolation in RdmDU degrades the prediction accuracy, we then conduct an empirical study in Section \ref{sec:ablation_freq} and find that the loss of high-frequency components of original images leads to the drop in the classification accuracy. Based on the findings, we propose to replace the bicubic with a detail-enhancement model, so that the RdmDU module can reconstruct an image with rich high-frequency details and retain the CNN's accuracy at a high value (above 90\%). Finally, we use a deep learning-based super-resolution model, namely the EDSR \cite{lim2017enhanced} network, as the upsampling operation. The experimental results show that RdmDU-SR can preserve 93.5\% natural accuracy, but also achieves consistently better adversarial robustness.

\begin{table}[h]
\caption{Evaluation of RdmDU modules on the robustness and prediction accuracy on STL10 dataset (Refer to Section \ref{sec:exp_setting} for the setting of attack methods).}
    \centering
    \scalebox{1}{
    \begin{tabular}{c|c|cccc}
         & \textbf{Accuracy} & \multicolumn{3}{c}{ \textbf{Robustness}} \\
    \midrule
    \textbf{STL10} &   Clean & FGSM& C\&W&Deep Fool \\
            \hline
        CNN-only & 1.000 &0.074&0.000&0.000\\
        w/ RdmDU-Bicubic &0.812 &0.485&0.625&0.539 \\
        w/ RdmDU-SR &0.935&0.565&0.742&0.664
    \end{tabular}
    }
    \label{tab:RdmDU}
\end{table}

\subsection{What is Lacking in Preserving Classification Accuracy?} \label{sec:ablation_freq}

To discover why the bicubic interpolation leads to a severe drop in natural accuracy, we first compare the original input and the output images of RdmDU-Bicubic. From Figure \ref{fig:bic_compare}, we see that visually, the difference, if any, between a certain image $x$ and its processed version $x^{\downarrow \uparrow}$ is imperceptible. However, 
if we transform the image into the frequency domain using the 2D-Fast Fourier Transform (FFT) $\gF(\cdot)$, we can see that the RdmDU-Bicubic tends to suppress the high-frequency components of the original image. Therefore, we hypothesize that the accuracy drop arises because of the loss of high-frequency components. To corroborate this hypothesis, we then conduct experiments to study the contribution of high-frequency components to the accuracy of a well-trained CNN model.

\textbf{Setup:} Given a CNN classifier, we remove an increasing amount of high frequency components from input images and examine how the prediction accuracy changes with respect to the loss of high-frequency components. Specifically, we preprocess input images in two steps, namely low-pass filtering and energy normalizing: 

{a).} The low-pass filter removes the high frequency components above a certain threshold. We first compute the spectrum $z$ of an input image $x$ via the 2D-FFT, i.e., $z=\gF(x)$. Next, we remove the components with frequencies above a given threshold $r$, and obtain the resultant spectrum $z'$, where 
\begin{equation}
	z'(\mu,\nu) = z(\mu, \nu) \cdot \mathbbm{1}\left(\frac{d((\mu,\nu),(c_h,c_w))}{\nicefrac{1}{2} \sqrt{h^2+w^2}} \leq  r \right).
\end{equation}
Here $d(\cdot ,\cdot )$ denotes the Euclidean distance between two points and $(c_h,c_w)$ is the position of the center, which corresponds to the zero-frequency component. The larger the value of $r$ is, the more high-frequency components are retained. 

{b).} Removing high-frequency components reduces the energy of the spectrum $E(z)$, where 
\begin{equation}
    E(z) \equiv \int_{\mu} \int_{\nu}|z(\mu, \nu)|^2 \mathrm{d}\mu \mathrm{d}\nu.
\end{equation}
To ensure that the comparisons are fair, we normalize the spectrums of the images before and after processing so that they have the same energy. Thus, we multiply the resultant spectrum $z'$ with $\sqrt{E(z)/E(z')}$ and reconstruct the RGB image for making subsequent predictions.

\def \SubFigWidth {0.16} 
\def \SubImgWidth {0.8}
\begin{figure*}[t!]
	\centering
	\begin{subfigure}{\SubFigWidth \linewidth}
		\centering
    	\includegraphics[width=\SubImgWidth \linewidth]{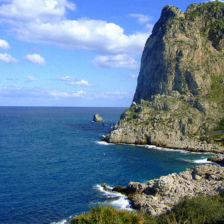}
	\end{subfigure}
	\begin{subfigure}{\SubFigWidth \linewidth}
		\centering
    	\includegraphics[width=\SubImgWidth \linewidth]{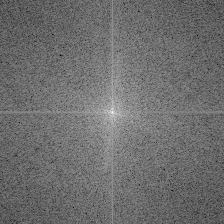}
	\end{subfigure}
	\begin{subfigure}{\SubFigWidth \linewidth}
		\centering
    	\includegraphics[width=\SubImgWidth \linewidth]{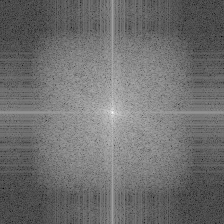}
	\end{subfigure}
	\begin{subfigure}{\SubFigWidth \linewidth}
		\centering
    	\includegraphics[width=\SubImgWidth \linewidth]{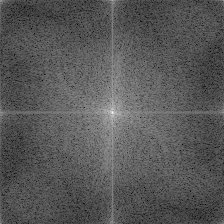}
	\end{subfigure}	
	\begin{subfigure}{\SubFigWidth \linewidth}
		\centering
    	\includegraphics[width=\SubImgWidth \linewidth]{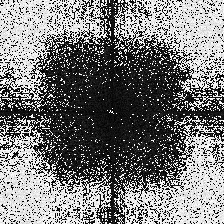}
	\end{subfigure}
	\begin{subfigure}{\SubFigWidth \linewidth}
		\centering
    	\includegraphics[width=\SubImgWidth \linewidth]{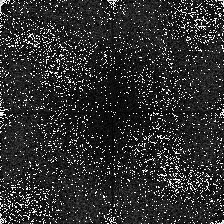}
	\end{subfigure}
	
	\vspace{1em}

	\begin{subfigure}{\SubFigWidth \linewidth}
		\centering
    	\includegraphics[width=\SubImgWidth \linewidth]{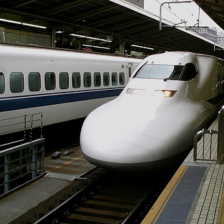}
	\end{subfigure}
	\begin{subfigure}{\SubFigWidth \linewidth}
		\centering
    	\includegraphics[width=\SubImgWidth \linewidth]{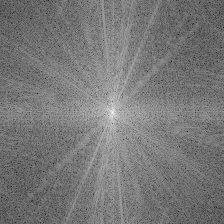}
	\end{subfigure}
	\begin{subfigure}{\SubFigWidth \linewidth}
		\centering
    	\includegraphics[width=\SubImgWidth \linewidth]{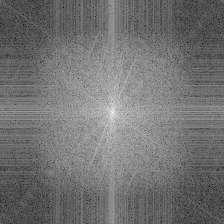}
	\end{subfigure}
	\begin{subfigure}{\SubFigWidth \linewidth}
		\centering
    	\includegraphics[width=\SubImgWidth \linewidth]{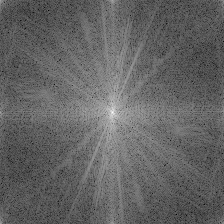}
	\end{subfigure}	
	\begin{subfigure}{\SubFigWidth \linewidth}
		\centering
    	\includegraphics[width=\SubImgWidth \linewidth]{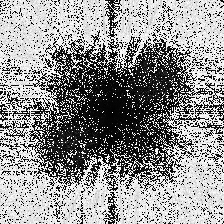}
	\end{subfigure}
	\begin{subfigure}{\SubFigWidth \linewidth}
		\centering
    	\includegraphics[width=\SubImgWidth \linewidth]{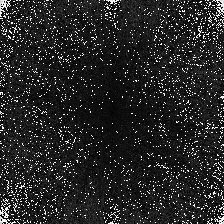}
	\end{subfigure}
	
	\vspace{1em}
	
	\begin{subfigure}{\SubFigWidth \linewidth}
		\centering
    	\includegraphics[width=\SubImgWidth \linewidth]{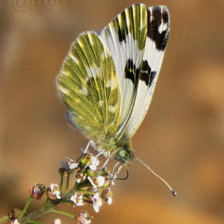}
    	\caption{$x$}
	\end{subfigure}
	\begin{subfigure}{\SubFigWidth \linewidth}
		\centering
    	\includegraphics[width=\SubImgWidth \linewidth]{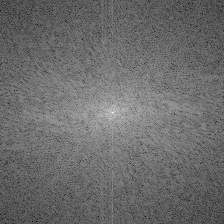}
    	\caption{$\mathcal{F}(x)$}
	\end{subfigure}
	\begin{subfigure}{\SubFigWidth \linewidth}
		\centering
    	\includegraphics[width=\SubImgWidth \linewidth]{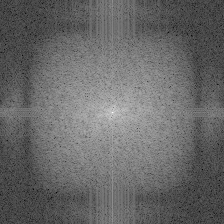}
    	\caption{$\mathcal{F}(x^{\downarrow \uparrow}_{\text{Bicubic}})$}
	\end{subfigure}
	\begin{subfigure}{\SubFigWidth \linewidth}
		\centering
    	\includegraphics[width=\SubImgWidth \linewidth]{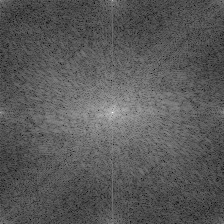}
    	\caption{$\mathcal{F}(x^{\downarrow \uparrow}_{\text{EDSR}})$}
	\end{subfigure}	
	\begin{subfigure}{\SubFigWidth \linewidth}
		\centering
    	\includegraphics[width=\SubImgWidth \linewidth]{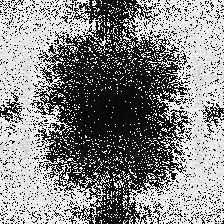}
    	\caption{$\mathcal{F}(x)-\mathcal{F}(x^{\downarrow \uparrow}_{\text{Bicubic}})$}
	\end{subfigure}
	\begin{subfigure}{\SubFigWidth \linewidth}
		\centering
    	\includegraphics[width=\SubImgWidth \linewidth]{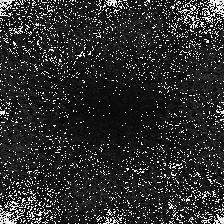}
    	\caption{$\mathcal{F}(x)-\mathcal{F}(x^{\downarrow \uparrow}_{\text{EDSR}})$}
	\end{subfigure}
	
    \caption{Comparison between images processed by RdmDU-Bicubic and RdmDU-SR. From left to right, column (a), original images $x$; column (b), spectrum of $x$; column (c), spectrums of images processed by RdmDU-Bicubic $\mathcal{F}(x^{\downarrow \uparrow}_{\text{Bicubic}})$; column (d), spectrums of images processed by RdmDU-SR $\mathcal{F}(x^{\downarrow \uparrow}_{\text{EDSR}})$; column (e), difference maps $\mathcal{F}(x)-\mathcal{F}(x^{\downarrow \uparrow}_{\text{Bicubic}})$; column (f), difference maps $\mathcal{F}(x)-\mathcal{F}(x^{\downarrow \uparrow}_{\text{EDSR}})$.}
    \label{fig:edsr_compare}
\end{figure*}

\begin{figure}[h]
    \centering
    \includegraphics[width=0.4\textwidth]{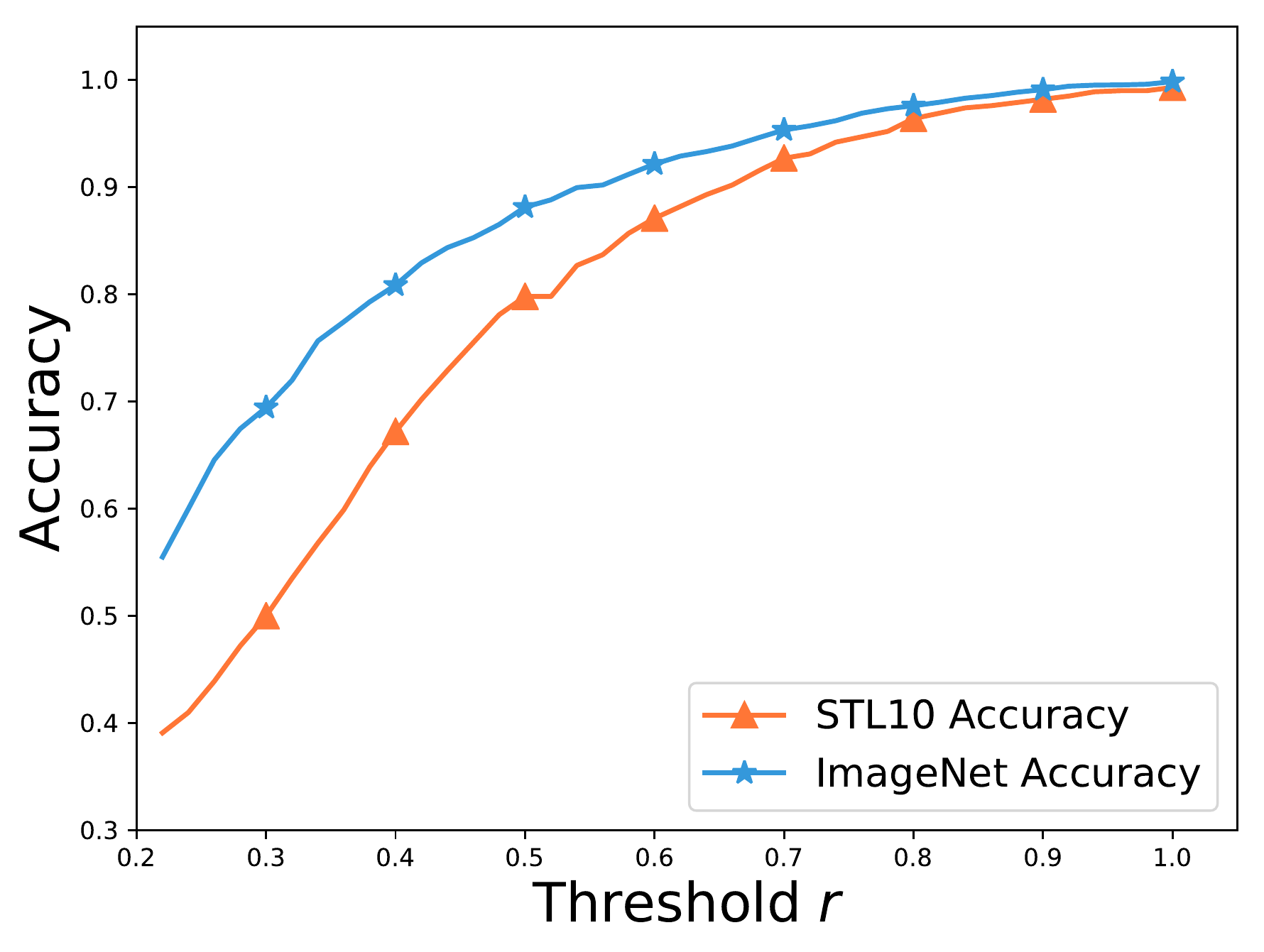}
    \caption{Accuracy v.s. the amount of high-frequency components removed. The smaller $r$ is, the more high-frequency components are removed. On both two datasets, we can see even the energy of the filtered spectrum is controlled to be the same with the origin, the loss of high-freq components still impairs the test accuracy.}
    \label{fig:fre}
\end{figure}

\textbf{Findings:} We conduct experiments on the STL10 and ImageNet datasets. For each dataset, we plot a curve of the performance of the classifier as a function of the threshold on the frequency $r$. From Figure \ref{fig:fre}, we find that, on the two datasets, the prediction accuracies drop rapidly as $r$ decreases, corresponding to an increase in the amount of high-frequency components being filtered out. Thus, the high-frequency components of the input images are critical for maintaining the accuracy of predictions. Given this observation, to preserve high natural accuracy, we propose to select an upsampling method that can super-resolve images while generating rich details. 

\textbf{RdmDU-SR:} Deep-learning-based SR methods have been shown to achieve impressive performances on SR tasks \cite{dong2015image, lim2017enhanced}. They are capable of generating high-frequency details such as rich textures and sharp edges. Here, we adopt the EDSR \cite{lim2017enhanced} model, which is among the most effective implementations for SR tasks and has been widely used as the backbone of various SR algorithms. We transform images processed by the EDSR into the frequency domain and compare the spectrums with those of images processed by the bicubic upsampling. The spectrums are shown in Figure \ref{fig:edsr_compare}, and we also calculate the heatmaps of the difference between the spectrums of processed images and original images. 

From Figure  \ref{fig:edsr_compare}, we can observe that, in each row, the four corners of $\mathcal{F}(x)-\mathcal{F}(x^{\downarrow \uparrow}_{\text{Bicubic}})$ are obviously bright, while the corners and the center of $\mathcal{F}(x)-\mathcal{F}(x^{\downarrow \uparrow}_{\text{EDSR}})$ are all relatively dark. It means that, in comparison to the bicubic interpolation, EDSR can effectively reconstruct high-freqency components of the original images. 
According to the previous findings, we can deduce that RdmDU-SR preserves the high natural accuracies of CNN classifiers because of its ability to generate rich details and textures.

\subsection{RAIN Framework} \label{sec:full_rain}

In the previous sections, we described two randomized preprocessing operations, both of which were shown to improve the robustness of CNN classifiers with minor degradations to the prediction accuracies. Here, we formally introduce our \emph{Robust and Accurate Image Classification} framework, which combines the RdmSCS and the RdmDU-SR. 
\begin{figure*}[t!]
	\centering
	\scalebox{1}{
		\includegraphics[width=1\linewidth]{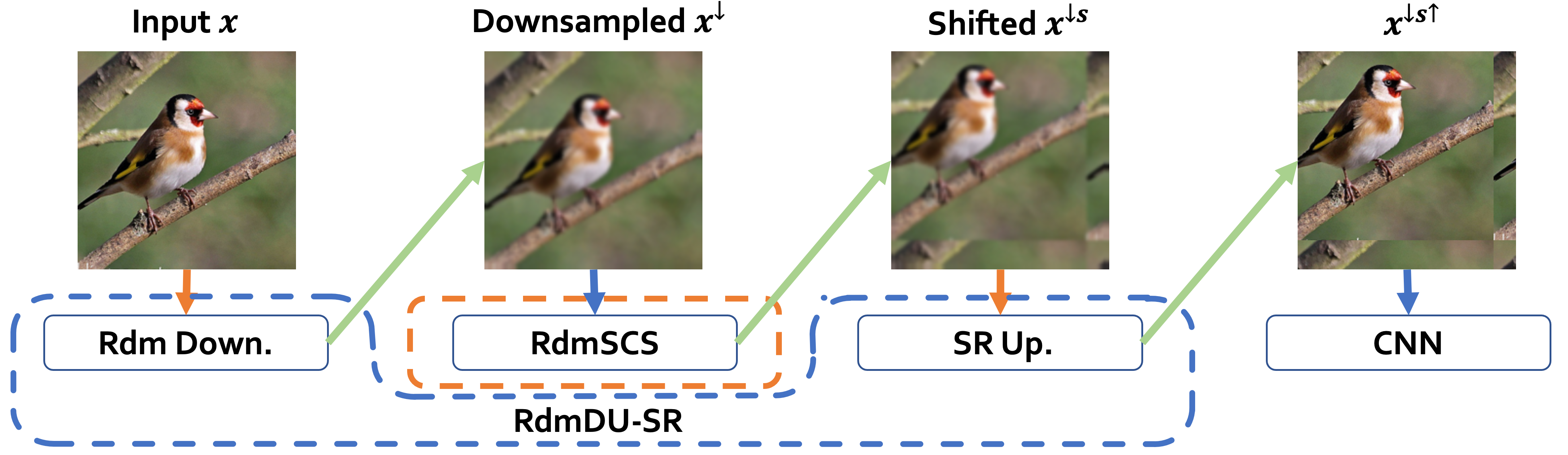}
	}
    \caption{The pipeline of our proposed RAIN framework: given an input, it is sequentially downsampled via RdmDU, circularly shifted via RdmSCS, and upsampled by an SR model within the RdmDU.}
    \label{fig:pipeline}
\end{figure*}

The whole pipeline of the RAIN is as follows (see Figure \ref{fig:pipeline}). For a given well-trained classification model $C(\cdot)$, RAIN first downsamples a certain image $x$ through the randomized downsampling operation within the RdmDU module and shifts the resultant image via the RdmSCS module successively. Then, a well-trained EDSR model upsamples the image so that the resulting size is the same as the original image, and the details are also enriched. Lastly, the resultant image is fed to the given CNN classifier $C(\cdot)$ for subsequent prediction. To generate image details that are most useful for $C(\cdot)$, we fix the CNN's parameters and fine-tune the EDSR model's parameters to minimize the loss of the classification task over a few training epochs. 

\section{Experiments}
\label{sec:exp}
In this section, we conduct extensive sets of experiments to evaluate the effectiveness of RAIN in terms of the enhancement of the CNNs' robustness and the preservation of their accuracies. We first examine how well the prediction accuracies of the CNN equipped with RAIN on clean images are retained, and evaluate the robustness against various white-box attacks and black-box attacks. Then, we also compare the proposed RAIN framework with other state-of-the-art methods to show the advantages of RAIN on robustness and accuracy. RAIN is a preprocessing-based method, we finally show that RAIN can work in conjunction with adversarial training. When dealing with the Expectation over Transformation (EoT) attacks, equipped with RAIN, the adversarially trained CNN classifiers achieve better natural accuracies. 

\subsection{Experimental Setting}
\label{sec:exp_setting}
We use {five} white-box attack methods \cite{goodfellow2014explaining, madry2017towards, moosavi2016deepfool, carlini2017towards, chen2018ead} and {three} black-box methods \cite{chen2017zoo, ilyas2018black, chen2020rays} to evaluate the effectiveness of the proposed RAIN. Since experiments are interspersed throughout Section \ref{sec:exp} and \ref{sec:ablation}, here we describe the common experimental settings. 

\subsubsection{Models and Datasets:} We conduct experiments on the STL10 \cite{coates2011analysis} and / or the ImageNet \cite{deng2009imagenet} datasets. We use a seven-layer CNN classifier for the STL10 dataset and the Resnet-50 \cite{he2016deep} model for the ImageNet. Details about the network architecures can be found in Section \ref{sec:arch}. For each dataset, we randomly select $5,000$ images from the test set for robustness evaluation, i.e., $M=5,000$ in Equation (\ref{eq:robustness_metric}). Besides, the performance of defense methods on original nature images is also evaluated on $\mathcal{T}$.

\subsubsection{White-box Attack Setting:} For the white-box attacks, we consider one single-step attack method (FGSM \cite{goodfellow2014explaining}) and four iterative attack methods (PGD \cite{madry2017towards}, C\&W \cite{carlini2017towards}, Deepfool \cite{moosavi2016deepfool} and EAD \cite{chen2018ead}). The FGSM and Deepfool attacks are based on the ${l_{\infty}}$ norm with $\epsilon = \nicefrac{8}{255}$. The PGD attack is based on the ${l_{\infty}}$ norm with $\epsilon = \nicefrac{8}{255}$ and learning rate $\nicefrac{2}{255}$. The C\&W attack is based on ${l_{2}}$ norm with learning rate $\nicefrac{1}{255}$. The EAD attack is implemented in EN-rule with learning rate $\nicefrac{1}{255}$ and $\beta=0.01$. All the four iterative attacks work over $40$ iterations of gradient descent. 

\subsubsection{Black-box Attack Setting:} For the black-box attacks, we consider the ZOO method \cite{chen2017zoo} with a maximum of 10,240 queries, the NES \cite{ilyas2018black} with a maximum of 2,000 queries, and the RayS \cite{chen2020rays} with a maximum 10,000 queries. All of these methods are based on the ${l_{\infty}}$ norm with $\epsilon = \nicefrac{8}{255}$. All the experiments follow the same settings discussed above unless otherwise specified.

\subsubsection{EoT-based Attack Setting:}
We evaluate the adversarial robustness against FGSM, PGD, and C\&W attacks under the EoT framework, in which each type of attack estimates the expected worst perturbation by averaging the obtained adversaries over a fixed number of queries. For the FGSM attack, the perturbation budget $\epsilon$ is $\nicefrac{8}{255}$ based on the $l_{\infty}$ norm; The C\&W attack is based on $l_2$ norm with rate $\nicefrac{1}{255}$; The PGD-1 attack based on $l_{\infty}$ with $\epsilon=\nicefrac{8}{255}$ and rate $\nicefrac{2}{255}$. 
Both of the PGD-1 and C\&W work in 40 iterations. All of these three attacks query the classifier for 40 times. We also use the PGD attack in another setting: based on $l_{\infty}$, PGD-2 works in 10 iterations with $\epsilon=\nicefrac{8}{255}$ and rate $\nicefrac{2}{255}$. The number of queries is set to be 10. 


\subsection{Adversarial Robustness of RAIN against White- and Black- box attacks} 
\label{exp_result}

We evaluate the robustness of our proposed RAIN on both of the STL10 \cite{coates2011analysis} and ImageNet \cite{deng2009imagenet} datasets under various white-box and black-box attacks. The experimental results demonstrate that the proposed RAIN framework improves the adversarial robustness of given CNNs, and, at the same time, manages to preserve high natural accuracies (>90\% on both the two datasets). 

We also compare RAIN with other four recent adversarial defense methods: two of them are preprocessing-based defense methods, namely Random Resizing \& Padding (RandR\&P) \cite{xie2017mitigating} and Pixel Deflection (PixelDef) \cite{prakash2018deflecting}; the other two are adversarial training-based  methods, namely Madry \etal \cite{madry2017towards} and Feature Denoising (FeatDn) \cite{xie2019feature} (both are trained with PGD examples of $\epsilon=\nicefrac{16}{255}$, step $=\nicefrac{1}{255}$ and 40 iterations).

\textbf{White-box Attack:} From Table \ref{tab:whitebox_result}, we observe that RAIN enhances the adversarial robustness of given CNNs, from 0\% to 78.1\% on STL10 against the C\&W attack and from 0\% to 86.7\% on ImageNet against the Deep Fool attack. In comparison to the two preprocessing-based competitors, RAIN achieves better adversarial robustness (e.g., 52.3\% v.s. 43.8\% against C\&W attack on ImageNet) but also maintains higher natural accuracies (e.g., 92.9\% v.s. 90.7\% of RandP\&R on STL10). For the two AT-based methods, RAIN also shows remarkable advantages regarding both robustness and natural accuracy. In specific, the accuraies of RAIN on both two datasets are at least 20 points higher than those of the two AT-based methods (e.g., 92.9\% v.s. 70.5\% on the STL10).

\begin{table*}[t!]
    \centering
    \caption{Robustness comparison of different defense methods against white-box attacks. We report the classification accuracies on clean images and various types of adversarial examples.  The results show that the proposed RAIN framework achieves the best performance in prediction accuracy and robustness. The methods of \cite{madry2017towards} and \cite{xie2019feature} train classifers with PGD adversarial examples. Thus, we do not evaluate their robustness with respect to PGD attacks for comparison.}
    \scalebox{1}{
    \begin{tabular}{l|c|ccccc}
    \multicolumn{1}{c|}{  } & \textbf{Accuracy} & \multicolumn{5}{c}{\textbf{Robustness}} \\
    \midrule
    \multicolumn{1}{c|}{ \textbf{STL10} }   &Clean      &FGSM   &PGD    &C\&W   &Deep Fool  &EAD\\
    \hline
    CNN-only                                & 1.000         &0.074  &0      &0      & 0         & 0\\
    RAIN                                        &\textbf{0.929} &\textbf{0.681} &\textbf{0.045} &\textbf{0.781} &\textbf{0.734} &\textbf{0.828}\\
    PixelDef \cite{prakash2018deflecting} & 0.883       & 0.397              & 0             &0.320          &0.117          &0.407 \\
     RandP\&R \cite{xie2017mitigating}  &0.907          &0.376          &0.025          &0.699          &0.305          &0.789\\
    Madry \cite{madry2017towards}               &0.705          & 0.607         &-              &0.289          &0.234          &0.028 \\
    FeatDn \cite{xie2019feature}     &0.696          &0.615          & -             &0.300          &0.227          &0.034\\

    \midrule
    \multicolumn{1}{c|}{ \textbf{ImageNet} } &Clean      &FGSM   &PGD    &C\&W   &Deep Fool  &EAD\\
    \hline
    CNN-only                                    &1.000      &0.125  &0      &0      &0          &0 \\
    RAIN                &\textbf{0.933} &\textbf{0.625} &\textbf{0.012} &\textbf{0.523} &\textbf{0.867}&\textbf{0.857} \\
    PixelDef \cite{prakash2018deflecting} & 0.858   & 0.357     &0  &0.320  &0.210  &0.365  \\
     RandP\&R  \cite{xie2017mitigating}     &0.928  & 343       &0  &0.438& 0.703   &0.759\\
    Madry \cite{madry2017towards}                   & 0.623 &0.598      &-  &0.355  &0.426  &0.063\\
    FeatDn \cite{xie2019feature}         & 0.653 &0.603      & - &0.398  &0.410  &0.115\\
    \end{tabular}
    }
    \label{tab:whitebox_result}
\end{table*}

\textbf{Black-box Attacks:} We compare the proposed RAIN framework to the baseline methods under black-box adversarial attacks.  Table \ref{tab:Black-box Result} shows that RAIN framework achieves the best black-box adversarial robustness against the ZOO and RayS attacks (91.2\% on STL10 and 88.5\% on ImageNet against ZOO, 88.2\% on STL10 and 92.4\% on ImageNet against RayS). In terms of the NES attack, the performance of RAIN is very close to that of the Random Resizing \& Padding method \cite{xie2017mitigating} (87.1\% v.s. 87.3\% on STL10 and 88.2\% v.s. 88.1\% on ImageNet), and is clearly better than the other three baselines competitors.

 \begin{table}[h!]
    \centering
    \caption{Evaluation of the robustness of the proposed RAIN framework and the baseline methods under black-box attacks on STL10 and ImageNet datasets.}
    \scalebox{1}{
    \begin{tabular}{l|c|cccc}
    \multicolumn{1}{c|}{  } & \textbf{Accuracy} & \multicolumn{2}{c}{ \textbf{Robustness}} \\
    \midrule
    \multicolumn{1}{c|}{ \textbf{STL10} }            & Clean    &ZOO    &NES    &RayS   \\
    \hline
    CNN-only                                        & 1         &0      &0.02   &0      \\
    RAIN                                            &\textbf{0.929}     & \textbf{0.912}    &0.871      & \textbf{0.882}\\
    PixelDef \cite{prakash2018deflecting}   &0.883              &0.679              &0.650      & 0.743\\
     RandP\&R \cite{xie2017mitigating}      &0.907              &0.854              &\textbf{0.873} & 828\\
    Madry \cite{madry2017towards}                   &0.705              &0.705              &0.663      &0.265\\
    FeatDn \cite{xie2019feature}         &0.696              &0.621              &0.594      &0.304\\

    \midrule
    \multicolumn{1}{c|}{ \textbf{ImageNet} }        & Clean    &ZOO    &NES    &RayS \\
    \hline
    CNN-only                                        & 1         &0      &0.02   &0      \\
    RAIN                                        & \textbf{0.933}    &\textbf{0.885}     &\textbf{0.882} & \textbf{0.924}\\
    PixelDef \cite{prakash2018deflecting} &0.858            &0.846              &0.841          & 0.848\\
    RandP\&R \cite{xie2017mitigating}  &0.928              &0.867              &0.881          & 0.913\\
    Madry \cite{madry2017towards}               &0.623              &0.620              &0.611          & 0.396\\
    FeatDn \cite{xie2019feature}     &0.653              &0.663           &0.641         & 0.421\\
    
    \end{tabular}
    }
    
    \label{tab:Black-box Result}
\end{table}{}

\subsection{Working in Conjunction with AT against EoT-based Attacks} \label{sec:eot_attack}
Defense methods \cite{xie2017mitigating, prakash2018deflecting, samangouei2018defense, dziugaite2016study} based on gradient obfuscation can block the back-propagation path of gradients from the predictions to inputs. Consequently, the attacker cannot accurately approximate the gradients to generate adversarial perturbations. 
However, Athalye \etal \cite{athalye2018obfuscated} demonstrated that attackers can circumvent the gradient obfuscation problem through the
Expectation over Transformation (EoT) technique for randomization methods or the Backward Pass Differentiable Approximate (BPDA) technique for non-differentiable transformations. Since our RAIN framework is built with two randomized transformations, RdmSCS and RdmDU, here we will discuss the effectiveness of RAIN against EoT-based attacks. 

Given an input $x$, it is firstly preprocessed by some random transformation $T$ before fed into the classifier $C(\cdot)$. Assume the random transformation $T$ is i.i.d. sampled from the distribution of transformation $P_{T}$, the EoT-based attack calculates the gradients of the expected prediction, $\nabla_x \E_{T\sim P_{T}} [C(T(x))]$. Since
\begin{equation}
    \nabla_x \E_{T\sim P_T} [C(T(x))] = \E_{T\sim P_{T}} [\nabla_x C(T(x))],
\end{equation}
the gradients of the expected prediction can be estimated by averaging the gradients of predictions of $x$ over multiple queries. When the number of query times is sufficiently large, the randomization-based defense will fail \cite{athalye2018obfuscated}. 

\textbf{RAIN-AT:} To deal with the EoT-based attack, it is imperative to perform adversarial training (AT) \cite{madry2017towards} on CNN classifiers. Within the RAIN framework, an input image is first modified with the RdmSCS and RdmDU-SR modules arranged in sequence; then, the processed image is sent to the \emph{adversarially trained CNN} classifier for making a prediction. To achieve the best performance, we then fine-tune the parametric EDSR model \emph{only} with natural clean samples (denoted as RAIN-AT). 

Conventional adversarial training suffers from the undesired trade-off between the prediction accuracy on natural images and the robustness against adversarial attacks \cite{zhang2019theoretically}. Equipped with our RAIN framework, we find that the adversarially trained CNN classifiers achieve higher natural accuracies but with no degradation to the robustness. Besides, since the pre-trained EDSR only needs to be fine-tuned with one epoch (on both the STL10 dataset and the ImageNet dataset), the implementation of RAIN does not incur too much extra computational cost.

\begin{table*}[t!]
    \centering
    \caption{ Robustness comparison of different defense methods against EoT-based attacks. We report the classification accuracies on clean images and various types of adversarial examples. The results show that the proposed RAIN framework with adversarially trained CNNs (RAIN-AT) achieves the best robustness against various attacks.}
    \scalebox{1}{
    \begin{tabular}{l|c|ccccc}
    \multicolumn{1}{c|}{ }      & \textbf{Accuracy} & \multicolumn{4}{c}{\textbf{Robustness}} \\
    \midrule
    \multicolumn{1}{c|}{ \textbf{STL10} }           &Clean  &FGSM   &PGD-1    &PGD-2  &C\&W \\
    \hline
    PixelDef \cite{prakash2018deflecting} 	&0.883  &0.082  & 0     & 0     & 0.283   \\
    RandP\&R \cite{xie2017mitigating} 		&0.907  &0.125  & 0.015 &0.024  &0.365\\
    RAIN											&\textbf{0.929}  &0.188	&0.01   &0.023  &0.402\\
    \hdashline
    Madry \cite{madry2017towards} 					&0.705  &0.607	&\textbf{0.367}  &0.363  &0.294\\
    RAIN-AT 										& 0.745 &\textbf{0.647}  &0.361  &\textbf{0.375}  &\textbf{0.605}\\
    RandP\&R-AT                                &0.619 &0.575  &0.364 &0.374   &0.428 \\

    \midrule
    \multicolumn{1}{c|}{ \textbf{ImageNet} }        &Clean  &FGSM   &PGD-1    &PGD-2  &C\&W \\
    \hline
    PixelDef \cite{prakash2018deflecting} 	&0.858  &0.022  & 0     & 0     & 0.247 \\
    RandP\&R  \cite{xie2017mitigating} 	    &0.928  &0.067	&0      &0      &0.276\\
    RAIN 											&\textbf{0.933}  &0.125	&0      &0      &0.255\\
    \hdashline
    Madry \cite{madry2017towards} 					&0.623  &0.598	&0.361  &0.361  &0.349\\
    RAIN-AT 									    &0.656	&\textbf{0.630}  &\textbf{0.363}	&\textbf{0.377}   &\textbf{0.473} \\
    RandP\&R-AT                                 &0.569 &0.574 &0.347 &0.356   &0.430
    \end{tabular}
    }
    \label{tab:eot}
\end{table*}

\textbf{Evaluation:} 
From Table \ref{tab:eot}, we observe that, based on normally trained CNN classifiers, RAIN and the other two preprocessing-based methods, namely PixelDef \cite{prakash2018deflecting} and RandP\&R \cite{xie2017mitigating}, are totally misled by EoT-based attacks (The accuracy is almost equal to 0\% against PGD attacks). In contrast, the adversarially trained CNNs \cite{madry2017towards} can effectively defend EoT-based attacks. More specifically,  Madry \etal \cite{madry2017towards} achieves 36.7\% accuracy against the PGD attack on the STL10 dataset and 36.1\% against PGD attack on the ImageNet dataset.

Based on the adversarially trained CNNs (Madry \cite{madry2017towards}), the RAIN framework further enhances adversarial robustness and improves the prediction accuracy on natural images. More precisely, RAIN-AT achieves 65.6\% natural accuracy against 62.3\% of Madry \cite{madry2017towards} on the ImageNet dataset and 74.5\% against 70.5\% on the STL10 dataset. The Random Padding \& Resizing \cite{xie2017mitigating} method also can be combined with adversarially trained CNNs (RandP\&R-AT). We can see that our RAIN-AT outperforms RandP\&R-AT consistently on all types of attacks. Unlike the dedicated preprocessing modules in our RAIN, the random padding and resizing operations results in a severe drop of accuracy (around 10 points lower than ours on both two datasets).

In summary, our RAIN and the adversarial training can work together to defend EoT-based attacks. Equipped with RAIN, the adversarially trained CNN can achieve better accuracy on clean images.

\section{Abaltaion Study}
\label{sec:ablation}
\subsection{On the Combination Order of RdmSCS and RdmDU} \label{sec:ablation_order}
The two randomized preprocessing modules can be combined in three different orders: RdmSCS+RdmDU, RdmDU+RdmSCS, and RAIN. Here, we conduct experiments on the STL10 dataset to show why we choose the RAIN framework with the modules arranged as in Figure \ref{fig:pipeline} instead of the other two variants. The experimental settings have been described in Section \ref{sec:exp_setting}. 

From Table \ref{tab:combinations_result}, we see that all of the three variations are clearly more robust than the original CNN model. In comparison to the results in Table \ref{tab:RdmSCS} and Table \ref{tab:RdmDU}, the combination of RdmSCS and RdmDU-SR clearly outperforms the separate utilization of these two modules for enhancing robustness. In real-world applications, since the probability that input images are naturally clean is much higher than that of adversarial examples, we choose the order of `Downsampling-Shifting-Upsampling', i.e., RAIN, which achieves the highest natural accuracy (92.9\%) among the three variants.

\begin{table}[h!]
    \centering
    \caption{Comparison of the three type of combinations of RdmSCS and RdmDU-SR (with $p=0.05$). 
    Among the three types,  the type RAIN achieves the best natural accuracy.}
    \scalebox{1}{
    \begin{tabular}{c|c|ccc}
     & \textbf{Accuracy} & \multicolumn{3}{c}{ \textbf{Robustness}} \\
    \midrule
    \textbf{STL10}  &Clean          &FGSM   &DeepFool   & C\&W \\
    \hline
    CNN-only 		& 1.000         &0.074  &0          &0\\
    RdmSCS-RdmDU  		&0.909          & 0.675	& 0.737     &0.843\\
    RdmDU-RdmSCS		&0.899          & 0.666	& 0.768     &0.797\\
    RAIN  		&\textbf{0.929} & 0.680 & 0.734     &0.781 

    \end{tabular}
    }
    \label{tab:combinations_result}
\end{table}{}

\subsection{On the Selection the Value of $p$} \label{sec:ablation_p}
In the proposed RAIN framework, the value of $p$ is an important hyperparameter; this parameter affects the robustness as well as the accuracy on natural images. We conduct experiments on the STL10 dataset to examine the effectiveness of RdmSCS with different values of $p$.

From Table \ref{tab:pvalue} and Table \ref{tab:RdmSCS}, we observe that, for each value of $p$, the combination of RdmSCS and RdmDU-SR (RAIN) results in significantly better adversarial robustness in comparison to using RdmSCS only. However, introducing an extra preprocessing module also deteriorates the natural accuracy on clean images. To preserve a high natural accuracy (above 90\%), we should set the value of $p$ to be less or equal to 0.1. Regarding the adversarial robustness, we choose $p=0.05$, which achieves the best robustness against FGSM and Deep Fool attacks and also maintains a 92.9\% natural accuracy.

\begin{table}[h]
\caption{Evaluation of RAIN with various values of $p$ on the robustness and prediction accuracy on STL10 dataset.}
    \centering
    \scalebox{1}{
      \begin{tabular}{c|c|cccc}
           & \textbf{Accuracy} & \multicolumn{3}{c}{ \textbf{Robustness}} \\
    \midrule
    \textbf{STL10}      &Clean  & FGSM  &DeepFool  &C\&W \\
            \hline
	    $p = 0.01$	    & 0.938	&0.623	        &0.695              &0.726 \\
        $p=0.05$	    & 0.929	&\textbf{0.683}	&\textbf{0.734}     &0.781 \\
        $p=0.1$	        & 0.909	&0.64	        &0.677              &0.820 \\
        $p=0.15$	    & 0.774	&0.59	        &0.703              &0.671
    \end{tabular}
    }
    \label{tab:pvalue}
\end{table}

\subsection{Effects of Adversarial Noise on Feature Maps}
Adversarial attacks add imperceptible noise to input images and results in a catastrophic failure of CNN classifiers. Even though the magnitude of perturbations on input images are small, the negative effects will be amplified across the hidden layers. In CNN classifiers, features generated at the last convolutional layer can reflect the semantic information and indicate pixels that are critical for subsequent prediction \cite{xie2019feature, mustafa2019image}. To investigate the efficacy of our RAIN framework, we extract and compare the feature maps of clean images and their adversarial examples.

We train a ResNet50 model on the ImageNet dataset and collect two images from the ImageNet for illustration. We use FGSM attack with $\epsilon=\nicefrac{8}{255}$ to generate the adversarial examples, $x^{\text C}_{\epsilon}$, of a certain input $x$. Here, the superscript `$\text C$' means the target model is the CNN only. 
We extract the features of $x$ and $x^{\text C}_{\epsilon}$ in the last convolutional layer for comparison. From Figure \ref{fig:feat_maps}, we see that the feature maps $F(x)$ focus on the region corresponding to the class objects in the original images (e.g., in the second row, the highlighted pixels corresponds to the position of airplane). We subtract the feature map of the adversarial example $x^{\text C}_{\epsilon}$ from that of $x$ and obtain the difference map $F(x^{\text C}_{\epsilon})-F(x)$. It is obvious that the adversarial feature map highlights many other irrelevant pixels in addition to the critical ones, which could lead to wrong predictions downstream. When the CNN classifier is equipped with RAIN, the adversarial examples are crafted in an end-to-end manner, i.e., the gradients of prediction are backpropagated through the defense modules. We denote the adversarial example in the RAIN framework as $x^{\text R}_{\epsilon}$. From the difference map $F(x^{\text R}_{\epsilon})-F(x)$ in Figure \ref{fig:feat_maps}, we observe that the feature map of $x^{\text R}_{\epsilon}$ is close to the feature map of clean image $x$. In comparison to results in column (\textbf{c}), the RAIN framework manages to suppress the malicious effects of the adversarial noise in the hidden layer. 

\def \SubFigWidth {0.24} 
\def \SubImgWidth {0.95}
\begin{figure}[h]
	\begin{subfigure}{\SubFigWidth\linewidth}
		\centering
    	\includegraphics[height=\SubImgWidth \linewidth]{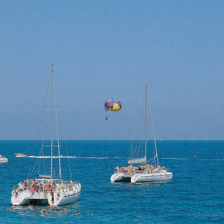}
	\end{subfigure}
	\begin{subfigure}{\SubFigWidth\linewidth}
		\centering
    	\includegraphics[height=\SubImgWidth \linewidth]{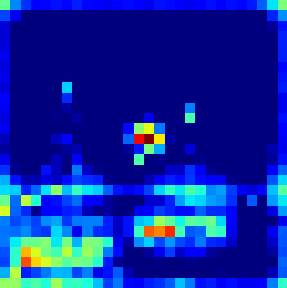}
	\end{subfigure}
	\begin{subfigure}{\SubFigWidth\linewidth}
		\centering
    	\includegraphics[height=\SubImgWidth \linewidth]{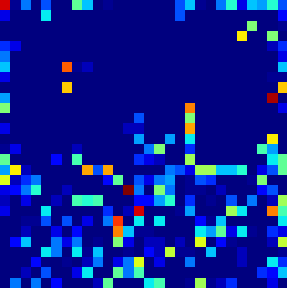}
	\end{subfigure}
	\begin{subfigure}{\SubFigWidth\linewidth}
		\centering
    	\includegraphics[height=\SubImgWidth \linewidth]{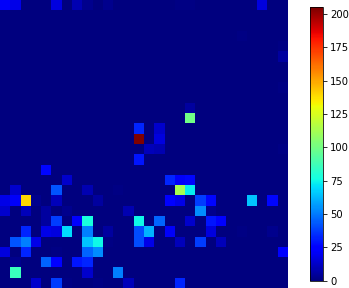}
	\end{subfigure}
	
	\vspace{1em}
	
	\begin{subfigure}{\SubFigWidth\linewidth}
		\centering
    	\includegraphics[height=\SubImgWidth \linewidth]{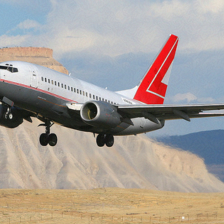}
    	\caption{\footnotesize $x$}
	\end{subfigure}
	\begin{subfigure}{\SubFigWidth\linewidth}
		\centering
    	\includegraphics[height=\SubImgWidth \linewidth]{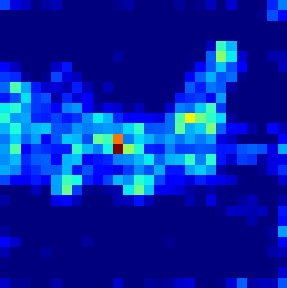}
    	\caption{\footnotesize $F(x)$}
	\end{subfigure}
	\begin{subfigure}{\SubFigWidth\linewidth}
		\centering
    	\includegraphics[height=\SubImgWidth \linewidth]{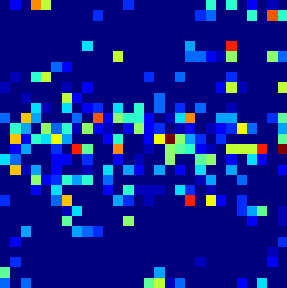}
    	\caption{\footnotesize $F(x^{\text{C}}_{\epsilon})-F(x)$}
	\end{subfigure}
	\begin{subfigure}{\SubFigWidth\linewidth}
		\centering
    	\includegraphics[height=\SubImgWidth \linewidth]{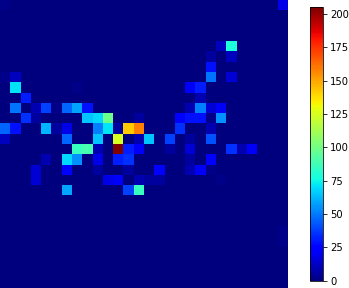}
    	\caption{\footnotesize $F(x^{\text{R}}_{\epsilon})-F(x)$}
	\end{subfigure}
    \caption{A comparison on the feature maps of clean images and their adversarial examples. In each row, column \textbf{(a)} shows the  natural image $x$; In column \textbf{(b)}, $F(x)$ denotes the feature map of $x$ extracted from the given CNN; column \textbf{(c)} shows the difference between the feature map of $x$ and that of the adversarial example $x^{\text C}_{\epsilon}$ generated w.r.t. the CNN; column \textbf{(d)} shows the difference between the feature map of $x$ and that of the adversarial example $x^{\text R}_{\epsilon}$ crafted w.r.t. the CNN equipped with our RAIN.}
    \label{fig:feat_maps}
\end{figure}

\section{Conclusions} \label{sec:conclusion}
This paper proposes a novel adversarial defense framework, RAIN, which combines two randomization modules, namely RdmSCS and RdmDU. As shown through our extensive experiments on two benchmark datasets, ,  RAIN can significantly enhance the robustness of given CNN classifiers and simultaneously preserve their high classification accuracies. RAIN preserves the reliability of CNN classifiers when the set of inputs is a combination of clean images and their adversarially perturbed versions.

In the future, we aim to develop a defense framework that is applicable to the classification of low-resolution and low-quality images because the quality of such images can be severely degraded by preprocessing operations, such as downsampling. The degradation may be so severe that effective SR models cannot reconstruct images to a level that predictions can be reliably made. We also plan to analyze the fundamental limits of the trade-off between prediction accuracy and adversarial robustness in a theoretically-grounded manner. This may lead to the proposal of a defense mechanism against adversarial attacks that has some theoretical guarantees.

\section*{Appendix}
\subsection{Networks Used on the STL10 and ImageNet datasets}
\label{sec:arch}
 
\textbf{CNN Classifier on the STL10:}
The CNN architecture used on the STL10 dataset is shown in Table \ref{tab:archi_stl10}. Our implementation builds on the open-source codes.\footnote{\url{https://github.com/aaron-xichen/pytorch-playground}}

 \begin{table}[h]
    \centering
    \caption{The architecture of the CNN classifier on the STL10 dataset. The classifier consists of either convolutional (Conv) layers or fully connected (FC) layers.} 
    \scalebox{1}{
    \begin{tabular}{c|c|c}

    \multicolumn{3}{c}{} \\
    \textbf{STL10}  &  & Layer \\
    \hline
    \multirow{5}{*}{Conv Layers} & $\times$1 & Conv(3, 32, 3, 1) + MaxPooling2d + ReLU \\
    & $\times$1 & Conv(32, 64, 3, 1) + MaxPooling2d + ReLU \\
    & $\times$1 & Conv(64, 128, 3, 1) + MaxPooling2d + ReLU \\
        & $\times$1 & Conv(128, 128, 3, 1) + MaxPooling2d + ReLU \\
            & $\times$2& Conv(128, 256, 3, 0) +  ReLU \\
    \midrule
    FC Layer & $\times$1 & MaxPooling2d + Linear(256,10)\\
    \multicolumn{3}{c}{} \\
    \end{tabular}
    }
    \label{tab:archi_stl10}
\end{table}

\textbf{CNN Classifier on the ImageNet:}
We use the Resnet-50 \cite{he2016deep} as the classification model on the Imagenet dataset. The pretrained version of the Resnet-50 can be found from the Torchvision repository.\footnote{\url{https://github.com/pytorch/vision/blob/master/torchvision/models}}

\textbf{EDSR:}
Our RAIN uses the same EDSR model for both of the two datasets. We use the single-scale baseline version of EDSR,\footnote{\url{https://github.com/thstkdgus35/EDSR-PyTorch}} which consists of only 16 residual blocks (see Table 1 in \cite{lim2017enhanced}).


\bibliographystyle{IEEEtran.bst}
\bibliography{egbib.bib}

%
%
%

\end{document}